\documentclass[prd,nofootinbib,twocolumn,showpacs]{revtex4}

\usepackage{graphics}
\usepackage{bm}

\def\be{\begin{equation}}
\def\ee{\end{equation}}
\def\ba{\begin{eqnarray}}
\def\ea{\end{eqnarray}}

\begin{document}

\title{Early reionization by cosmic strings revisited}

\author{Levon Pogosian and Alexander Vilenkin}
\affiliation{Institute of Cosmology, Department of Physics and Astronomy,
Tufts University, Medford, MA 02155, USA}

\begin{abstract}
Measurements of the CMB temperature anisotropy and the
temperature-polarization cross correlation by WMAP
suggest a reionization redshift of $z \sim 17 \pm 5$.
On the other hand, observations of high redshift galaxies
indicate a presence of a significant fraction of neutral hydrogen
at redshift $z\sim 6-7$. 
We show that cosmic strings with tensions
well within, but not far from, current observation bounds could cause
early star formation at a level sufficient to explain the high reionization
redshift.
\end{abstract}


\

\maketitle
Most of the baryonic matter in the present day universe is in the
form of ionized plasma. Yet we know that the universe was neutral immediately
after the recombination. Therefore, sometime between now and
redshift $z\sim 10^3$ the universe was reionized.
The detailed history of reionization is a mystery and has become a
particularly hot topic of discussion following the
release of WMAP's first year data. The WMAP team has reported
an optical depth to the last scattering $\tau = 0.17 \pm 0.06$, which
for the instantaneous and homogeneous reionization implies
a redshift of $z\sim 17\pm 5$ \cite{wmap_spergel,wmap_kogut}.
The precise value of the reionization redshift inferred from CMB data
depends on the assumed ionization history and a more conservative interpretation
of WMAP's result is that the measured optical depth to last scattering is
consistent with a significant ionized fraction
at redshifts $z\sim 11-30$ \cite{wmap_kogut}.

The commonly accepted general picture of reionization is that it was
caused by first stars and quasars emitting high energy photons.
In order for the universe to be significantly ionized by
redshift $z_r \sim 20$ a sufficiently large fraction of matter would need
to be collapsed to form stars prior to $z_r$.
It remains to be seen if collapse of Gaussian initial
inhomogeneities can produce the required number of reionizing stars formed by such
early time \cite{LoebBarkana04}.

To add to the puzzle, the WMAP value of the re-ionization redshift is significantly
higher than the value of $z_r \sim 6-7$ which one would infer from measurements of
Lyman-$\alpha$ absorption in the spectra of high redshift galaxies
(e.g. see \cite{bunker04}).
This could mean that the ionized fraction evolved non-monotonically
with redshift and involved processes more complex than previously thought
\cite{WL03,Cen03,WL04,LBH04,gnedin04}.

In this paper we investigate the possibility that early reionization
was triggered by stars formed in the wakes of moving cosmic strings.
The idea that strings could cause reionization
has been discussed in a number of papers
\cite{rees86,hara93,avelino03}.  Refs. \cite{rees86,hara93} were
written over a decade ago and used cosmological parameters that
were significantly different from the ones we know now. Also, at
the time those papers were written, cosmic strings were considered
a prime candidate to explain the formation of large scale
structure and were generally attributed a higher value of mass per
unit length then currently allowed by observations.  In
Ref.~\cite{avelino03} stings are represented by a spectrum of
linear perturbations. It is not clear to us why this should give
an adequate description, since we know that the inhomogeneities
induced by strings on scales relevant to reionization are highly
non-linear in character \footnote{This is confirmed by numerical
studies of cosmic string wakes \cite{robert97}}. Here, we
re-investigate the reionization by strings in the light of the
current data, and with nonlinear string dynamics fully taken into
account.

Another motivation for this study is the recent realization that
cosmic strings (or D-branes) are naturally produced in many brane
inflation scenarios, motivated by String/M Theory
\cite{costring,jst2,PTWW03,DV03,Polchinski03}.  In these theories
strings can move and interact in extra dimensions, in addition to the
observed three spatial dimensions. In particular, while appearing to
intersect in our three dimensions, they may actually miss each other
in the extra dimension(s). Hence, the effective intercommutation rate
of these strings will generally be lower than one.  As a consequence,
one would expect more strings per horizon in these theories, with more
small scale structure accumulating on the strings \cite{DV03}. This
can have interesting observational consequences which must be
investigated.

Let us now consider the
problem in more quantitative detail. The approximate fraction of total baryonic
matter that needs to be in stars in order to ionize the rest of the gas
can be roughly estimated by accounting for the following main effects.
The ionization energy of hydrogen is $13.6$ eV, while nuclear fusion
in stars produces $\sim 7$ MeV per hydrogen atom. Not all photons emitted
by stars have energies above $13.6$ eV and each hydrogen atom can recombine
more than once. The number of ionizing photons
produced per baryon depends on the mass and the composition of the
star which produced them. Massive, metal-free stars (the so-called
Population III start) can produce roughly $10^5$ ionizing photons per baryon.
Locally observed stars, which generically have a much lower mass
and some metal content, can produce $\sim 4000$ ionizing photons
per baryon \cite{WL03}. Only a small fraction of the gas in the halo
will have time to form Population III stars, since
those will soon explode as supernovae and prevent the formation of other
supermassive stars. Hence, star formation regions
quickly become dominated by low-mass stars.
The fraction of matter that would need to be in stars in order
to reionize the entire universe can roughly be estimated \cite{bromm00,bromm03,WL03}
to be
\be
f_{\rm stars} \sim 10^{-3}-10^{-4} (\eta/10) \ ,
\label{neededf}
\ee
where $\eta$ (typically $>1$ for $z>7$) is the number of ionizing photons
needed per hydrogen atom. In (\ref{neededf}) the lower number corresponds
to metal free stars. However, the metallicity is likely to be increased rather
quickly as the first stars explode as supernovae, hence we are going
to use a more conservative estimate of $f_{\rm stars}$, corresponding
to the higher number in (\ref{neededf}).

What if WMAP's estimates are true and the universe was indeed
reionized around redshift $z\sim 20$? Assuming that cosmic structures
grew by gravitational instability from initial Gaussian density
fluctuations, one could apply the so-called extended version of the
Press-Schechter model \cite{Bond91} and estimate the fraction of
matter that collapsed and formed stars prior to $z\sim 20$.  Some
analytical and numerical work estimating that fraction has recently
been done in \cite{LoebBarkana04}. The fraction of matter in stars is
related to the quantity $F_G(M_{\rm min})$, which is the fraction of
matter in halos of mass $M_{\rm min}$ or greater, where $M_{\rm min}$
is the minimum mass that a halo must have in order to form a
galaxy. For $M_{\rm min} \approx 7 \times 10^{5} M_{\odot}$ Barkana
and Loeb find $F_G(M_{\rm min}) \approx 4 \times 10^{-5}$
\cite{LoebBarkana04}, which is roughly consistent with the needed
fraction given in eq.~(\ref{neededf}).  However, the precise estimates
of $f$ using this method and the use of Press-Schecher itself at such
high redshift is still a matter of some debate.

We are going to suggest that cosmic strings could provide the fraction
which is just as large and hence play a significant role in the early
reionization.  We will discuss current observational bounds on cosmic
strings later in the paper, but only mention now that the strings we
consider have tensions that are consistent with the most recent
observations of CMB and large scale structure (LSS) and, therefore, do
not contribute appreciably to structure formation on large scales.

During the radiation and matter dominated eras
the string network evolves according to a scaling solution
\cite{BenBouch,AllenShellard90,AlTur89} which
on sufficiently large scales can be
described by two length scales. The first scale, $\xi(t)$, is the coherence
length of strings, i.~e. the distance
beyond which directions along the string are uncorrelated.
The second scale, $L(t)$, is the average interstring separation.
Scaling implies that both length scales grow in proportion to the horizon.
Cosmic string simulations indicate that $\xi (t) \sim t$,
while
\be
L(t) = \gamma t \ ,
\label{Lt}
\ee
with $\gamma \approx 0.8$ in the matter era \cite{BenBouch,AllenShellard90}.
The so-called one-scale model \cite{Kibble,shellard96}, in which the
two length scales are taken to be the same, has been
quite successful in describing the general properties of cosmic string networks
inferred from numerical simulations. These simulations have assumed that
cosmic strings would reconnect on every intersection. It is of interest
to us, however, to also consider the case when the reconnection probability
is less than one. Then, because of the straightening of wiggles on subhorizon
scales due to the expansion of the universe, one would still expect
$\xi(t) \sim t$, but the string density would increase, therefore reducing the
interstring distance. Hence, smaller intercommutation probabilities
imply smaller $\gamma$.

Numerical simulations show that long strings possess significant amounts of
small-scale structure in the form of kinks and wiggles on scales much smaller
than horizon. To an observer who cannot resolve this structure, the string
will appear to be smooth, but with a larger effective mass per unit length
${\tilde \mu}$ and a smaller effective tension ${\tilde T}$. An unperturbed
string (with $\mu=T$) exerts no gravitational force on nearby particles.
In contrast, a wiggly string with ${\tilde \mu} > {\tilde T}$ attracts
particles like a massive rod. The effective equation
of state of a wiggly strings is ${\tilde \mu} {\tilde T} = \mu^2$
\cite{Carter,AV90} and the
velocity boost given by a moving wiggly string to nearby matter is
\cite{VacVil91,Vollick92}
\be
u_i = 4\pi G {\tilde \mu} v_s \gamma_s
+ {2\pi G ({\tilde \mu}-{\tilde T})\over v_s \gamma_s } \ ,
\label{ui}
\ee
where $v_s$ is the string velocity and $\gamma_s=(1-v_s^2)^{-1/2}$.


Let us now consider a wake \cite{SV84} formed behind a string
segment of length $\xi(t)\sim t$ that travelled with a speed $v_s$
at some early time $t_i$ for a period of time $\sim t_i$. For now,
let us assume that $t_i \gtrsim t_{eq}$, where $t_{eq}$ is the
time of radiation-matter equality. There would be wakes formed
during the radiation era as well, and we will discuss them
separately later in the paper. The length $l_w(z)$, the width
$w_w(z)$ and the thickness $d_w(z)$ of the wake will evolve with
redshift as \cite{tanmay86,Stebbins92,VSbook} \ba \label{dwz} l_w
\sim t_i {z_i \over z} \ , \ w_w \sim v_s t_i {z_i \over z } \ , \
d_w \sim u_i t_i \left({z_i \over z }\right)^2 \ , \ea as long as
\be z > z_i u_i/v_s \ . \label{vscondition} \ee At smaller
redshifts the wake thickness $d_w$ becomes comparable to the width
$w_w$ and the wake takes on a shape of a cylinder whose diameter
grows as $z^{-3/2}$. We find that the condition
(\ref{vscondition}) is satisfied for all wakes that have a chance
to play a role in the early reionization. Numerical simulations
\cite{AllenShellard90} show that
average string velocities on scales comparable to the horizon are
of order $v_s \sim 0.15$. The effective mass per unit length of
these strings is ${\tilde \mu} \approx 1.6 \mu$, which implies that the
second term in eq.~(\ref{ui}) will dominate\footnote{Strings with
lower inter-commutation probabilities are expected to accumulated even
more small-scale structure due to the suppression of loop production \cite{DV03}.}:
\be
u_i \approx {2\pi G  \mu \alpha\over v_s} \ ,
\label{uiw}
\ee
where we have defined $\alpha \equiv ({\tilde \mu}-{\tilde T})/\mu$.

As the universe expands, the fraction of matter in the wakes grows in
proportion to the scale factor. For wakes formed at redshift $z_i$,
this fraction is
\be
f_w(z,z_i) \sim {l_w w_w d_w \over {\gamma^2 t_i^3 (z_i/z)^3}}
\sim 2\pi G \mu \alpha \gamma^{-2} {z_i \over z} \ ,
\ee
where we have used Eq.~(\ref{Lt}) for the average inter-string distance.
To make quantitative estimates easier, let us re-write this
using some characteristic values for the parameters:
\be
\label{f_w}
f_w(z,z_i) \sim 10^{-3} \gamma^{-2} \left({20 \over z}\right)
\left({z_i \over z_{eq}}\right)
\left({ G\mu \alpha \over 10^{-6} }\right) \ ,
\ee
where we have used $z_{eq}\approx 3400$. Whether any part of this
fraction collapses into luminous objects depends
on the values of $z_i$, $z$ and $G\mu\alpha$.

What are the conditions leading to the formation of luminous objects
within a wake? As it grows, the wake will fragment
(due to the wiggliness of the string that produced the wake and
due to intersections with smaller wakes produced at earlier times)
into chunks of size
comparable to the wake thickness $d_w$ \cite{hara93,AvelinoShellard95}.
These fragments form the CDM halos into which the baryons will
fall after decoupling from photons.
Generally, one would expect the baryonic gas inside the wake to
collapse when its Jeans length becomes smaller than the thickness of the
wake. The Jeans length is defined as
\be
L_J(z)= c_s(z) \left( \pi \over G\rho_{bw}(z) \right)^{1/2} \ ,
\label{ljz}
\ee
where $c_s(z)$ and $\rho_{bw}(z)$ are, respectively, the speed of sound and
the energy density of baryons inside the wake. The infall velocity
grows with time as \cite{VSbook}
\be
u(z) \approx {2 \over 5} u_i \left({z_i \over z}\right)^{1/2}  \ .
\label{uz}
\ee
After a certain time the infall becomes supersonic and shocks develop
on either side of the baryonic wake\footnote{One could allow for the possibility for the
Jeans condition to be satisfied before the infall into the wake becomes
supersonic. However, the condition for supersonic infall, $c_s \lesssim u(z)$, and
the Jeans condition, roughly $c_s t \lesssim d_w(z)$, are essentially the same
inequality. Hence, it is sufficient to consider only the wakes that
form shocks.}.
The compression of the gas
within the shock will heat it to a temperature \cite{LL87}
\be
T_{\rm shock} \approx {3 \over 16} m_H u(z)^2 \ ,
\label{tshock}
\ee
where $m_H$ is the hydrogen mass. This implies that the baryonic
speed of sound inside the wake, $c_s \sim \sqrt{2T_{\rm shock}/m_H}$, is
comparable to $u(z)$:
\be
\label{csz}
c_s(z) \approx 0.6 u(z) \ .
\ee

The density of baryons inside the shock is enhanced by a factor of 4 compared
to the background baryon density. In addition, if
the gas is able to cool to some equilibrium temperature $T_{\rm cool}$,
its density would be enhanced by an additional factor
$T_{\rm shock} / T_{\rm cool}$. Taking these effects into account,
the baryon density in the wake can be written as
\be
\rho_{bw}(z) \approx {4X_b \over 6\pi G t_i^2} {T_{\rm shock} \over T_{\rm cool}}
\left({z \over z_i}\right)^3 \ ,
\label{rhobwz}
\ee
where $X_b \equiv \Omega_b/\Omega_M$ is the fraction of matter in baryons.
Substituting eqns (\ref{uz}), (\ref{csz}) and (\ref{rhobwz}) into (\ref{ljz}) gives
\be
L_J(z) \sim X_b^{-1/2}
\left({T_{\rm cool} \over T_{\rm shock}}\right)^{1/2}
u_i t_i \left({z_i \over z}\right)^2  \ .
\ee
The collapse condition $d_w(z) \geq L_J(z)$ then leads to
\be
X_b^{-1/2} \left({T_{\rm cool} \over T_{\rm shock}}\right)^{1/2} \lesssim 1 \ ,
\ee
or,
\be
{T_{\rm shock} \over T_{\rm cool}} \gtrsim 6 \ ,
\label{cooling}
\ee
where we have used $X_b \approx 0.16$.
From eqns.~(\ref{tshock}), (\ref{uz}) and (\ref{uiw}) it follows that
\ba
T_{\rm shock} &\approx&
{3 \pi^2 m_H \over 25} \left({G \mu \alpha \over v_s} \right)^2 {z_i \over z}
\nonumber \\  &\approx& 600 {\rm K}  \ {z_i \over z} \
\left({G\mu\alpha \over 10^{-6} }\right)^2
\left({0.15 \over v_s}\right)^2 \ ,
\ea
The lowest temperature to which the gas can cool depends on its metal
abundance. The metal free primordial gas can cool via atomic transitions of
hydrogen and helium down to $10^4$K, while molecular hydrogen
can cool the gas down to $200$K \cite{BL01}. Once the gas is enriched with
metals, it could in principle cool down to the CMB temperature.
Since the metallicity is likely to increase quickly, as discussed earlier,
we are going to adopt
\be
T_{\rm cool}(z)=T_{cmb}(z) = 2.726 (1+z) {\rm K} \ .
\ee
The collapse condition in eq.~(\ref{cooling}) can now be written as
\be
{z_i^{(m)} \over z^2} \gtrsim 0.03
\left({ 10^{-6} \over G\mu\alpha }\right)^2 \left({v_s \over 0.15}\right)^2 \ ,
\label{after}
\ee
where the superscript $(m)$ on $z_i$ denotes matter era. This, in turn, implies
a constraint on the value of $G\mu$ (using $z_i^{(m)} \leq z_{eq} \sim 3400$):
\be
G\mu \gtrsim  0.6 \times 10^{-7} \alpha^{-1} \left({z \over 20}\right)
\left({v_s \over 0.15}\right) \ .
\label{gmucool}
\ee
Formation of gaseous objects can only occur after the recombination,
because of the Compton drag. We are mainly interested
in the formation at much later redshifts, with ineq.~(\ref{gmucool}) providing
the necessary condition.

So far, we have only discussed wakes formed {\it after} $t_{eq}$. Let
us now consider wakes formed at some time $t_i$ during the radiation era.
For these wakes
the gravitational instability only sets in at $t\sim t_{eq}$. At earlier
times, the surface density of the wakes decreases as $t^{-1/2}$,
while the fraction of dark matter accreted onto all wakes formed within
a Hubble time of $t_i$ remains roughly constant \cite{tanmay86}.
The infall velocity into these wakes
will decrease with time as $(1+z)$ until the matter starts to
dominate. Therefore, for a wake formed at some $t_i<t_{eq}$, the infall
velocity at $t>t_{eq}$ will be given by
\be
{\tilde u}(z) \approx {2 \over 5} u_i \left({z_{eq} \over z}\right)^{1/2}
\left({z_{eq} \over z_i}\right) \ .
\ee
Consequently, the temperature inside the shocks that would form
in such wakes at $t>t_{\rm dec}$ would be given by
\be
{\tilde T}_{\rm shock}(z,z_i) \approx
{ 3 \pi^2 m_H \over 25}\left({G \mu \alpha \over v_s}\right)^2
{z_{eq}^3 \over z \ z_i^2}
\ee
The necessary condition given by eq.~(\ref{cooling}) will
now become
\be
z_i^{(r)} \lesssim  55000 \left({20 \over z}\right)
\left({G\mu \alpha \over 10^{-6} }\right) \left({0.15 \over v_s}\right) \ .
\label{before}
\ee
Again, we can use $z_i^{(r)}>z_{eq}$ to re-write this as a lower bound on
the value of $G\mu$, which leads to the same constraint as in eq.~(\ref{gmucool}).

Ultimately, we want to find the fraction of matter in all wakes, formed before
and after $t_{eq}$, that satisfy the collapse condition in eq.~(\ref{cooling}).
This total fraction can be expressed as
$f_w(z,z_i)$ from eq.~(\ref{f_w}) evaluated at $z_i=z_{eq}$ and multiplied by a
factor $a_w$. This factor can be written as a sum of the contributions from
wakes formed before and after $t_{eq}$:
\be
a_w=a_w^{(r)}+a_w^{(m)} \ .
\ee
Because the fraction of matter in wakes formed within each Hubble time (defined
as the time required for the horizon to double in size) remains roughly
constant during the radiation era, $a_w^{(r)}$ is simply equal to the
number of Hubble times between $z_{eq}$ and $z_i^{(r)}$, which is
bounded by inequality (\ref{before}). Namely,
\ba
a_w^{(r)} &\approx& {\rm log}_2 \left( {z_i^{(r)} \over z_{eq}} \right)^2
\nonumber \\
&\lesssim& 8 + 2 {\rm log}_2 \left[ \left({20 \over z}\right)
\left({G\mu \alpha \over 10^{-6} }\right) \left({0.15 \over v_s}\right)
\right]
\ .
\ea
The fraction of matter in wakes formed after $t_{eq}$ decreases with the
redshift as $z_i^{-1}$. Therefore, the dependence of $a_w^{(m)}$ on the number
of Hubble times between $z_{eq}$ and $z_{i}^{(m)}$ is sufficiently weak
and we can simply set $a_w^{(m)} \sim 1$.
Hence, the total fraction of matter in the wakes that satisfy the
collapse condition can be written as
\be
f_w(z) \sim 10^{-3} \gamma^{-2} a_w \left({20 \over z}\right)
\left({ G\mu \alpha \over 10^{-6} }\right) \ ,
\ee
where
\be
a_w \approx 9 + 2 {\rm log}_2 \left[ \left({20 \over z}\right)
\left({G\mu \alpha \over 10^{-6} }\right) \left({0.15 \over v_s}\right)\right] \ .
\label{aw}
\ee
Note that $a_w$ decreases
with a decrease in $G\mu$ or an increase in $z$, e.~g. $a_w \sim 10$ for
$G\mu\alpha \sim 10^{-6}$ and $z\sim 20$ yet $a_w \sim 1 $ for
$G\mu\alpha \sim 10^{-7}$ and $z\sim 30$.

The condition in eq.~(\ref{cooling})
does not say anything about the ability of the gas to cool. Only some fraction
$f_{\rm s}$
of the total number of baryons satisfying this condition will actually
be able to cool and form
stars. At present, there is no good theory of this fraction.
However, one could assume, based on the current ratio of the average mass density
in stars to the total baryon density, that this fraction is of order $10$\%
\cite{FHP98}.
Finally, we can estimate the fraction of matter in stars, formed in the wakes
behind cosmic strings, as
\be
f_{\rm w} \sim
10^{-4} {\gamma^{-2}} a_w \left({20 \over z}\right)
\left({f_{\rm s}\over 0.1}\right)
\left({G\mu\alpha \over 10^{-6} }\right)
\label{final}
\ee
This, together with eq.~(\ref{neededf}) and the WMAP's upper bound on the
reionization redshift ($z_r \lesssim 30$), can be used to put an approximate
upper bound on the value of $G\mu$:
\be
G \mu \lesssim 10^{-6} \left({\gamma^2 \alpha^{-1} a_w^{-1} \over 0.1} \right)
\left({\eta \over 10}\right) \left({0.1 \over f_{\rm s}}\right) \ ,
\label{upperbound}
\ee
which is roughly an order of magnitude higher than the upper bound reported
in \cite{avelino03}.
Inequalities (\ref{upperbound}) and (\ref{gmucool}) define the approximate
range of values of $G\mu$ for which strings may play a role in the
early reionization.
Interestingly, this range is quite narrow -- only an order of magnitude
wide. However, this range also depends on parameters $\alpha$ and $\gamma$,
which could be significantly different from
their canonical values if, e.~g., strings were D-branes formed at the end of
brane inflation \cite{costring,jst2,PTWW03,DV03,Polchinski03}.

Current CMB and LSS data imply that
$G\mu \lesssim 10^{-6}$ \cite{PTWW03,shellard03,PWW04}\footnote{These constraints
only apply to local cosmic strings considered in this paper. For recent
constraints on global strings see \cite{Bevis04}.}.
This bound, however, also depends on the number of strings within
a horizon or, more generally, on the scaling parameter $\gamma$.
When this dependence is taken into account, the constraint based
on CMB and LSS power spectra becomes
\be
G\mu \lesssim 10^{-6} \gamma \ .
\label{gmucmb}
\ee
Substituting this into eq.~(\ref{final}) gives
\be
f_{\rm w} \lesssim 10^{-3} \left({\gamma^{-1} a_w \alpha \over 10}\right)
\left({20 \over z}\right)
\left({f_{\rm s}\over 0.1}\right)  \ ,
\label{final1}
\ee
which is of the same order of magnitude as the fraction needed to
reionize the universe.


It may be worth mentioning
that recently there has been a detection of
galaxy lensing that looks very
much like a lensing by a cosmic string \cite{sazhin}. It remains to
be confirmed, but assuming it is a cosmic string, the implied value
of the string tension would be $G\mu \sim 4 \times 10^{-7}$. For this value
of $G\mu$, and assuming that $\alpha \approx 1$, from eq.~(\ref{aw}) it follows
that $a_w \approx 7$ at $z \sim 20$.
Recent analysis of the WMAP and SDSS data \cite{PWW04} shows that data
may actually prefer a non-zero contribution from strings, corresponding
to $G\mu \sim 4 \times 10^{-7} \gamma$. The combination of these
two measurements would imply ${\gamma \sim 1}$, which for
$z \sim 20$, would give
\be
f_{\rm w} \approx 3 \times 10^{-4} \left({f_{\rm s}\over 0.1}\right) \ ,
\label{optimist}
\ee
which is close to what would be required for the reionization by
redshift $z\sim 20$.

Our analysis was based on the expectation that the metallicity of the
gas in the wakes will quickly reach the level sufficient to prevent
the formation of heavy metal-free stars (the so-called Population 
III stars). If, however, a significant
number of Population III stars managed to form prior to metal enrichment,
then our results would be modified. In particular, a smaller
fraction of matter in the wakes would satisfy the collapse condition
(\ref{cooling})
due to the higher value of $T_{\rm cool}$ for a metal free gas.
On the other hand, a smaller value of $f_{rm star}$ would be required 
to reionize the universe. In combination, this is likely to reduce the
upper bound on $G\mu$ (eq.~(\ref{upperbound})) and 
increase the threashold value in eq.~(\ref{gmucool}), thus, 
narrowing the range of string parameters relevant for the reionization.
If a sizable fraction of matter was ionized by Population III 
stars in the wakes it could lead to a non-monotonic reionization similar
to the two-step reionization models of \cite{WL03,cen03}.

In conclusion, our analysis has shown that cosmic strings can play a
role in early reionization, provided that their mass per unit length
is within the range given by inequalities (\ref{gmucool}) and
(\ref{upperbound}). This range is consistent with, and may even be
somewhat favored by the current data.


\acknowledgments
We thank Avi Loeb for very helpful comments upon reading the draft of
this paper and Ken Olum for useful discussions.

\end{document}